\documentclass[aps,pre,floatfix]{revtex4}

  \usepackage[pdftex]{graphicx,color}

\newcommand{\beq}{\begin{equation}}
\newcommand{\eeq}{\end{equation}}
\newcommand{\be}{\begin{equation}}
\newcommand{\ee}{\end{equation}}
\newcommand{\bea}{\begin{eqnarray}}
\newcommand{\eea}{\end{eqnarray}}
\newcommand{\bean}{\begin{eqnarray*}}
\newcommand{\eean}{\end{eqnarray*}}
\newcommand{\ep}{\epsilon}
\newcommand{\e}{{\rm e}}
\renewcommand{\i}{{\rm i}}
\newcommand{\tsfrac}[2]{{\textstyle\frac{#1}{#2}}}
\newcommand{\deriv}[2]{{\frac{{\rm d} #1}{{\rm d} #2}}}
\def\prt#1#2{{\pa #1 \over \pa #2}}
\def\prtt#1#2{{\pa^2 #1 \over \pa #2 ^2}}
\newcommand{\pa}{\partial}

\begin{document}

\title{Pattern formation in the damped Nikolaevskiy equation}
\author{S. M. Cox}\author{P. C. Matthews}
\affiliation
{School of Mathematical Sciences, 
University of Nottingham,\\
University Park, Nottingham NG7 2RD, UK}

\begin{abstract}

The Nikolaevskiy equation has been proposed as a model for seismic waves, 
electroconvection and weak turbulence; we show that it can also be used to 
model transverse instabilities of fronts.  This equation possesses a 
large-scale ``Goldstone'' mode that significantly influences the stability 
of spatially periodic steady solutions; indeed, all such solutions are 
unstable at onset, and the equation exhibits so-called soft-mode 
turbulence. In many applications, a weak damping of this neutral mode will 
be present, and we study the influence of this damping on solutions to the 
Nikolaevskiy equation.  We examine the transition to the usual Eckhaus 
instability as the damping of the large-scale mode is increased, through 
numerical calculation and weakly nonlinear analysis. The latter is 
accomplished using asymptotically consistent systems of coupled amplitude 
equations. We find that there is a critical value of the damping below 
which (for a given value of the supercriticality parameter) all periodic 
steady states are unstable. The last solutions to lose stability lie in a 
cusp close to the left-hand side of the marginal stability curve.

\end{abstract}
\maketitle

\section{Introduction}

The Nikolaevskiy equation
\be
\prt u t+u \prt u x  =
-\prtt{}x \left( r  u - \left(1+\prtt{}x\right)^2 u \right) 
\label{nik}
\ee
has been widely studied, because of its application to several
physical systems and its interesting nonlinear dynamics. 
It can also be written in the alternative form 
\be
\prt \phi t+ \frac 1 2  \left(\prt \phi x\right)^2  =
-\prtt{}x \left( r  \phi - \left(1+\prtt{}x\right)^2 \phi \right) 
\label{nik2}
\ee
after introducing $u=\phi_x$. 

The equation (\ref{nik}) was first derived, in an extended form, by 
Nikolaevskiy as a model for seismic waves in the earth's crust 
\cite{ber,nik}. More recently, other applications of (\ref{nik}) or 
(\ref{nik2}) have been proposed. Fujisaka and Yamada \cite{fy} and
Tanaka \cite{tan04} have derived 
(\ref{nik2}) as a possible phase equation arising in reaction--diffusion 
systems.  Turbulence in electroconvection was studied experimentally by 
Hidaka et al.~\cite{hia}, who proposed that the chaotic dynamics was of a 
type described qualitatively by (\ref{nik}). More generally, the 
Nikolaevskiy equation can be regarded as a canonical model for a 
pattern-forming system with the additional feature of symmetry under the 
transformation $\phi \to \phi + {}$constant, in the form (\ref{nik2}) 
\cite{trt}, or with Galilean symmetry, in the form (\ref{nik}) \cite{mat}. 
Therefore, (\ref{nik}) is a suitable model for pattern-forming systems 
with these additional symmetries. Finally, as we show in 
Section~\ref{sec:fronts} below, (\ref{nik2}) can be used to describe 
finite-wavelength instabilities of travelling fronts.

The Nikolaevskiy equation exemplifies what is known as ``soft-mode 
turbulence''.  This form of chaotic dynamics arises from the interaction 
between a pattern of finite wavenumber, which appears for $r>0$, and a 
long-wave neutral (or ``Goldstone'') mode. The neutral mode arises as a 
direct consequence of the additional symmetries discussed above. In a 
sufficiently large domain, all spatially periodic steady states are 
unstable, as was shown by Tribelsky and Velarde \cite{triv}, and numerical 
investigations of (\ref{nik}) show irregular chaotic patterns 
\cite{mat,trt,st07}.  In Ref. \cite{mat} we showed that in this turbulent 
regime, (\ref{nik}) exhibits an unusual scaling, with the amplitude of $u$ 
proportional to the $3/4$ power of the supercriticality parameter $r$, 
whereas it is more usually proportional to the $1/2$ power in other 
pattern-forming systems. Fujisaka et al.~\cite{fhy} extended (\ref{nik2}) 
to two spatial dimensions and found that this same scaling law holds.

In this paper, we consider the effect of adding a weak damping term to the 
neutral mode in (\ref{nik}). The motivation for this is that in real 
experimental systems, the symmetry that gives rise to the neutral mode 
will often be weakly broken.  For example, true Galilean symmetry is broken by 
the existence of distant boundaries. In the case of electroconvection, a 
weak magnetic field damps the neutral mode \cite{htk}.  A second 
motivation for including the damping term is to improve our understanding 
of the appearance of chaotic dynamics in the undamped equation 
(\ref{nik}).  If the neutral mode is strongly damped, then the dynamics of 
patterns is governed by the Ginzburg--Landau equation and stable, steady 
patterns are observed. As the damping is reduced, a transition to 
Nikolaevskiy chaos can be observed and investigated. Our work extends the 
recent study by Tribelsky~\cite{try06}.

In Sections~\ref{sec:fronts} and~\ref{sec:damped}, we motivate and discuss 
the undamped and damped versions of the Nikolaevskiy equation. Then in 
Section~\ref{sec:3} we compute numerically the steady spatially periodic 
``roll'' solutions of the damped equation and compute their secondary 
stability. This numerical calculation shows that there is an additional 
pocket of stable rolls, beyond those considered by Tribelsky~\cite{try06}; 
indeed this pocket is significant in containing the last rolls to remain 
stable as the damping coefficient is reduced. Most of the stability 
boundary of the rolls corresponds to large-scale modulational 
instabilities; we then analyse these instabilities in three regimes, 
depending on the relative sizes of the supercriticality and damping 
parameters. In the strong-damping regime, the relevant weakly nonlinear 
description is a modified Ginzburg--Landau equation.  In the two other 
regimes, where the damping is either moderate or weak, our analyses are 
based on three coupled amplitude equations, for the amplitude and phase of 
the rolls and for an associated large-scale field; in each case, these 
amplitude equations may be reduced to a single, third-order evolution 
equation for the phase. In all three regimes, we examine the secondary 
stability of rolls, and in certain cases we are also able to determine the 
effects of the leading nonlinear term and comment on the direction of the 
bifurcation. We demonstrate the subcritical onset of instability in some 
cases, including the no-damping case, consistent with the observed sudden 
onset of the instability in numerical simulations of (\ref{nik}).

\section{The Nikolaevskiy equation as a model for transverse instability 
of fronts}
\label{sec:fronts}

\begin{figure}
\begin{center}
\includegraphics[width=0.8\linewidth]{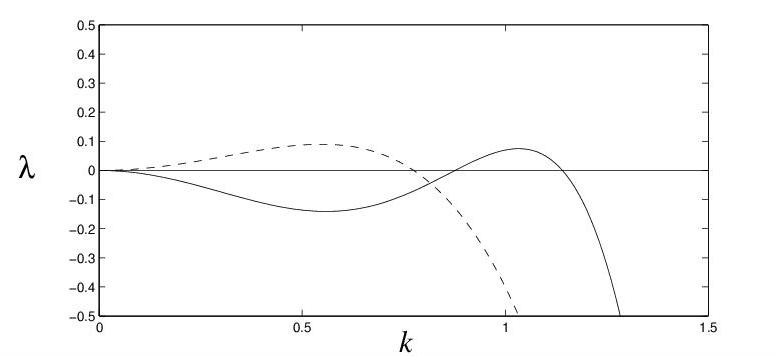}
\end{center}

\caption{Curves illustrating two possible dispersion relations for front 
instabilities. Solid line: finite-wavenumber instability, leading to 
Nikolaevskiy equation (type I). Dashed line: long-wave instability, 
leading to Kuramoto--Sivashinsky equation (type II).}

\label{fig:kurnik}
\end{figure}

In this section we explain how the Nikolaevskiy equation may arise as a 
model for finite-wavenumber transverse instabilities of planar fronts. 
Consider a planar travelling front in a homogeneous, isotropic medium, 
arising from, for example, a combustion problem or a system of 
reaction--diffusion equations. Suppose that the front is travelling at 
speed $c$ in the $z$ direction and that the $x$ coordinate is directed 
along the front. Consider now small perturbations to the uniform front. 
Let the position of the front under perturbation be denoted by $z = c t + 
\phi(x,t)$. Now the evolution equation for $\phi$ must have the property 
that it does depend on the value of $\phi$, because of translational 
invariance; the equation can depend only on $x$-derivatives of $\phi$.  
Furthermore, reflection symmetry in $x$ means that the only permissible 
linear terms involve even derivatives of $\phi$. Therefore, the linearized 
equation for $\phi$ must have the form
 \be
\prt \phi t = a_2  \frac{\pa^2 \phi}{\pa x^2} +a_4  \frac{\pa^4
  \phi}{\pa x^4} + a_6  \frac{\pa^6 \phi}{\pa x^6}+ \cdots,
\label{lineq}
 \ee
where the $a_j$ are constants. The corresponding dispersion relation is
 \be
\lambda = -a_2 k^2 + a_4 k^4 - a_6 k^6 + \cdots, \label{disp}
 \ee 
for the growth of modes $\phi \sim \exp(\lambda t + \i k x)$. From 
(\ref{disp}) it is apparent that there are two possible types of 
instability of the front. If $a_2 <0$ and $a_4 <0$, then the planar front 
is unstable to a band of wavenumbers near $k=0$ (Figure~\ref{fig:kurnik}, 
dashed line). This case has been widely studied, and leads, in the weakly 
nonlinear regime, to the Kuramoto--Sivashinsky equation 
\cite{kurbook,siva}, which, with suitable rescalings of $x$ and $t$, may 
be written in the form
 \be
\prt \phi t+ \frac 1 2  \left(\prt \phi x\right)^2  =
-\prtt{}x \left( r  \phi + \prtt{\phi}x\right).
\label{kseq}
 \ee 
But if $a_2$, $a_4$ and $a_6$ are all positive, then long-wave modes 
are stable and a finite-wavenumber instability is possible, as illustrated 
by the solid line in Figure~\ref{fig:kurnik}. After a rescaling of $x$ 
and $t$, (\ref{lineq}) then corresponds to the linear terms in 
(\ref{nik2}). These two possible types of instability are sometimes 
referred to as ``type II'' and ``type I'', respectively 
\cite{bansagi,cross}.

In either case, a nonlinear term must be added to the equation to limit 
the growth of the instability; its derivation is the same for either type 
of instability, and, as with the linear terms, only $x$-derivatives of 
$\phi$ may appear. The nonlinear term can be derived by a simple 
geometrical argument, considering the propagation of a tilted front 
\cite{kurbook}. Such an argument leads to a nonlinear term which in our 
scaling takes the form $(\pa \phi/\pa x)^2/2$, as in (\ref{nik2}) and 
(\ref{kseq}).

We have shown that (\ref{nik2}) may be regarded as a model equation for 
the transverse instability of a planar front at finite wavenumber. More 
generally, it is apparent that any physical system that gives rise to the 
Kuramoto--Sivashinsky equation might also lead to the Nikolaevskiy 
equation, depending on which of the two possible forms shown in 
Figure~\ref{fig:kurnik} is taken by the dispersion relation.
Another example~\cite{fy,tan04,tk} is the derivation of nonlinear
phase equations;   
this leads to the Nikolaevskiy and Kuramoto--Sivashinsky equations, in
different parameter regimes, as the dispersion relation changes from
type I to type II. The argument in the case of phase equations is
essentially identical to that given above for fronts, following from
the invariance under addition of a constant to the phase, and under
reflection in $x$.  

There are many studies of the transverse linear stability of fronts in the 
literature, arising from a variety of different physical and chemical 
systems. A review of the literature reveals several examples of a 
finite-wavenumber (type I) instability, consistent with the Nikolaevskiy 
equation. For example, in their study of an evaporation front for 
condensed matter heated by a laser, Anisimov et al.~\cite{ate} found a 
linear spectrum of type I for transverse instability. Both types of 
instability were found in an experimental and theoretical study of a 
reaction front between two chemicals in a Hele-Shaw 
cell~\cite{bansagi,dhkdw,kydw}.

\section{The damped Nikolaevskiy equation}
\label{sec:damped}

The complex chaotic behavior of the Nikolaevskiy equation at onset arises 
from the interaction between the unstable modes near $k=1$ and the neutral 
mode at $k=0$. However, in real physical systems it is likely that the 
symmetry giving rise to the neutral mode (such as translation or Galilean 
symmetry) will be weakly broken. For example, in the case of a propagating 
front, considered in Section~\ref{sec:fronts}, the translation symmetry 
may be broken by the presence of boundaries, or by a slight variation in 
the basic state with position. With such a weak symmetry breaking, the 
$k=0$ mode will not be truly neutral, but will be weakly damped instead.

A further motivation for including a damping term in the Nikolaevskiy 
equation is that it allows us to study the transition from ordered stable 
patterns to soft turbulence. Such a probing of the origins of the 
turbulent state is not possible in the original Nikolaevskiy equation 
(\ref{nik}), since chaos is observed directly at onset (provided the 
domain is sufficiently large). Introduction of a damping term gives 
another parameter, which can be used to control and investigate the onset 
of chaos. A similar approach has been used to examine the dynamics of the 
Kuramoto--Sivashinsky equation \cite{brunet,misbah}.

We thus consider in this paper the ``damped Nikolaevskiy'' equation
\be
\prt u t+u \prt u x  =
-\prtt{}x \left( r  u - \left(1+\prtt{}x\right)^2 u \right) 
- \nu\left(1+\prtt{}x\right)^2 u, \label{eq:damp}
 \ee 
with damping coefficient $\nu\geq0$; in particular we shall be concerned 
with the dynamics of (\ref{eq:damp}) near the onset of pattern formation. 
In examining the stability of roll solutions, we shall find it useful to 
consider various cases for the relative sizes of the supercriticality 
parameter
 \[
 r=\ep^2 
 \]
and the damping coefficient $\nu$, and hence we write
 \be 
\nu=\ep^s\mu, 
\label{eq:numu}
 \ee 
where $\mu=O(1)$; we shall examine below a variety of pertinent values 
for $s$.

Of course there are many ways in which one might add damping to the 
Nikolaevskiy equation. The damping term in (\ref{eq:damp}) is chosen so 
that all modes are linearly damped (with the exception of the mode with 
wavenumber $k=1$); large-scale modes decay at a rate $-\nu$. Furthermore, 
significant analytical simplifications follow from the fact that in 
(\ref{eq:damp}) the onset of linear instability is independent of $\nu$ 
(i.e., the critical value of $r$ is $r_c=0$ and the critical
wavenumber is $k_c=1$, for any $\nu\geq0$). Note that our choice of 
damping term differs from that of Tribelsky \cite{try06}, who considers 
instead a damping term proportional to $-u-\partial^2u/\partial x^2$. 
Consequently, in his formulation both $r_c$ and $k_c$ become functions of 
$\nu$, leading to additional algebraic complications.  However, the two 
formulations are equivalent under appropriate rescalings of $x$ and $t$, 
and appropriate mappings between the two parameter sets.

\section{Secondary stability of rolls: numerical results}
\label{sec:3}

We begin our investigation of (\ref{eq:damp}) by numerically computing 
spatially periodic steady ``roll'' solutions, and determining their 
secondary stability. To do so, first we fix $\ep$ and compute the roll 
solution $\bar{u}(x)$ for a given choice of $k$ and $\nu$ by integrating 
(\ref{eq:damp}) forwards in time in a box of length $2\pi/k$. In our 
pseudospectral numerical code to achieve this, $\bar{u}$ is approximated 
by the truncated Fourier series
 \[
\bar{u}(x)=\sum_{-N}^{N} \bar{u}_n\e^{n\i kx}.
 \] 
Disturbances $u'(x)$ to this roll solution then satisfy, from 
(\ref{eq:damp}),
 \be 
\prt{u'}t+u' \prt{\bar{u}}x +\bar{u}\prt{u'}x = 
-\prtt{}x \left( r u' - \left(1+\prtt{}x\right)^2 u' \right) - 
\nu\left(1+\prtt{}x\right)^2 u', 
\label{eq:dist}
 \ee 
and these disturbances are sought in the form 
 \be
u'(x)=\e^{\sigma t+\i p x}\sum_{-N}^{N} u_n'\e^{n\i kx} 
 \label{eq:uprime}
 \ee
(where the sum is again truncated for numerical purposes). Equation 
(\ref{eq:dist}) is thus reduced to a matrix eigenvalue problem for the 
growth rate $\sigma$. By computing $\sigma(p;k,\ep,\nu)$ for $p$ in the 
range $-k/2\leq p\leq k/2$, we are then able to determine the stability of 
the roll solution $\bar{u}(x)$.

\begin{figure}
\begin{center}
(a)\includegraphics[width=0.67\linewidth]{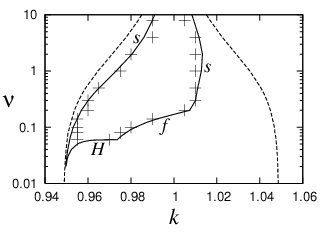}

(b)\includegraphics[width=0.67\linewidth]{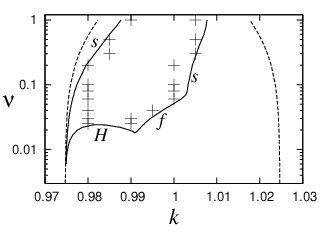}
\end{center}

\caption{Stability boundaries of rolls in (\ref{eq:damp}), computed 
numerically for (a) $\ep=0.1$ and (b) $\ep=0.05$. Rolls with wavenumber 
$k$ exist between the dashed curves, and are stable inside the region 
enclosed by solid curves. The annotations on the stability boundary 
indicate the type of instability suffered by the rolls upon crossing that 
part of the boundary: modulational steady (``\textit{s}''), modulational 
Hopf (``\textit{H}'') or finite-wavelength (``\textit{f}'') instability. 
The crosses correspond to numerical simulations of the initial-value 
problem (\ref{eq:damp}) in a large computational box and indicate the most 
extreme values of the wavenumber for which rolls are found to be stable to 
small perturbations (necessarily restricted to those that fit into the 
computational box); that all crosses lie inside the roll stability 
boundary provides a consistency check on our results.}

\label{fig:ep0.1}

\end{figure}

Results for $\ep=0.1$ and $\ep=0.05$ are summarized in 
Figure~\ref{fig:ep0.1}, where rolls are stable inside the regions bounded 
by solid curves. For large enough damping (i.e., towards the top of each 
plot), the familiar Eckhaus stability boundaries are recovered. For 
sufficiently small damping, by contrast, all rolls are unstable -- this 
shows that the known instability of all rolls in the undamped Nikolaevskiy 
equation extends to the case of (sufficiently small) finite damping, for a 
fixed value of the supercriticality parameter. Intermediate numerical 
results show the transition between the two limits.

Several features of Figure~\ref{fig:ep0.1} merit comment. First, the 
region of stable rolls, which is roughly symmetrical about $k=1$ at large 
damping, becomes distinctly asymmetrical as the damping is reduced; rolls 
with $k>1$ tend to be less stable than those with $k<1$. Second, stability 
is finally lost, as $\nu$ is decreased, in a small cusp near the left-hand 
side of the marginal stability curve. For $\ep=0.1$, this corner is at 
$\nu\approx0.02$ and $k\approx0.950$ (the roll existence boundary is, for 
this value of $\nu$, at $k\approx0.9493$); for $\ep=0.05$, we have found 
it much harder to compute the very thin cusp right down to its tip.  The 
left and right sides of the cusp correspond, respectively, to monotonic 
and oscillatory long-wavelength instabilities. In fact the entire boundary 
corresponds to long-wavelength instabilities ($|p|\ll1$), with the 
exception of the segment marked ``\textit{f}'', where the instability has 
finite wavelength ($p=O(1)$). For even smaller values of $\ep$ than shown 
in the figure, this finite-wavelength segment is absent, a point to which 
we shall return in Section~\ref{sec:5.3.1}. 

\begin{figure}
\begin{center}
\includegraphics[width=0.6\linewidth]{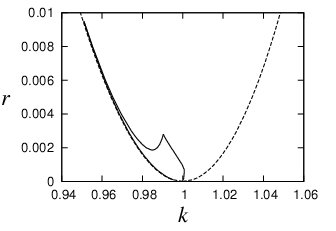}
\end{center}
\caption{Stability boundaries of rolls in (\ref{eq:damp}), computed
numerically for $\nu=0.02$. Rolls with wavenumber $k$ exist between
the dashed curves, and are stable inside the region enclosed by solid
curves.}
\label{fig:mu0.02}
\end{figure}

A more conventional picture of the stability balloon for rolls is shown in 
Figure~\ref{fig:mu0.02}, where the damping parameter $\nu$ is fixed to be 
$0.02$ and the stability of rolls is shown in the $(k,r)$-plane. Again 
rolls are stable inside the solid curves. The region of stable rolls lies 
almost entirely in the region $k<1$. For $r>0.003$, all rolls are 
unstable, except those exceedingly close to the left-hand part of the 
marginal stability curve; these rolls are last to lose stability as $r$ is 
increased.  Finally, for $r>0.01$, all rolls are unstable. A corresponding 
picture for smaller damping coefficient $\nu$ would show stable rolls in a 
similar, but smaller region, confined closer to the point of onset ($r=0$, 
$k=1$). In the next section, we shall argue on the basis of asymptotic 
arguments that in the limit $\nu\to0$ (approaching the undamped 
Nikolaevskiy equation), the size of the stable region shrinks to zero. 
Crucially, however, we expect a small, but finite, region of stable rolls 
in the $(k,r)$-plane for any nonzero value of the damping coefficient 
$\nu$. It is {\em only} for the undamped case $\nu=0$ that all rolls are 
unstable at onset.

\section{Secondary stability of rolls: analytical results}

The numerical results in the previous section reveal a number of different 
instabilities of rolls in the damped Nikolaevskiy equation. 
In this section, we consider in detail three scalings for $\nu$ 
in (\ref{eq:damp}) which shed particular light on the numerical stability 
results illustrated above: these correspond to choosing in turn $s=1$, 
$s=3/2$ and $s=2$ in (\ref{eq:numu}).

\subsection{Strong damping: $s=1$}
\label{sec:strong}

The case $s=1$ represents strong damping, in the sense that the decay rate 
$\nu = \ep \mu$ of the large-scale mode is $O(\ep)$, which is much
greater than the  
$O(\ep^2)$ growth rate of the pattern-forming mode. The effect of this 
strong damping is simply to modify the usual Ginzburg--Landau amplitude 
equation and thus modify the Eckhaus stability boundary, as we now 
demonstrate. Our conclusions below are consistent with those obtained by 
Tribelsky~\cite{try06}, 
although his analysis proceeds directly from a 
substitution of (\ref{eq:uprime}) into (\ref{eq:dist}), rather than the 
amplitude-equation framework developed below.

We apply to (\ref{eq:damp}) the usual weakly nonlinear scalings and expand 
the solution as
 \be
 u = \ep(A (X,T) \e^{\i x} + {\rm c.c.})  + \ep^2 u_2 + \ep^3 u_3+\cdots,
 \ee
where $X = \ep x $ and $T = \ep^2 t$, and where c.c.\ denotes the 
complex conjugate of the preceding term. At $O(\ep)$, (\ref{eq:damp}) is
satisfied. At $O(\ep^2)$, we find
 \[
 u_2=(-\tsfrac{1}{36}\i A^2\e^{2\i x} + {\rm c.c.})+f(X,T),
 \]
where the large-scale mode $f(X,T)$ is arbitrary. At $O(\ep^3)$, an
equation arises for the mean mode $f$, which represents a balance
between the driving term $(|A|^2)_X$ and the damping term $ - \mu f$:
\be
0=- (|A|^2)_X - \mu f.
\label{eq:felim}
\ee
Note that the linear terms $f_T$ and $f_{XX}$ do not appear at this order. 
The governing equation for $A$ is also obtained at $O(\ep^3)$, by applying
the usual solvability condition. It turns out to be
\be
A_T  =  A +  4A_{XX} -\tsfrac1{36} |A|^2 A  - \i f A,
\ee
from which $f$ may be eliminated using (\ref{eq:felim}) to give a single
amplitude equation for $A$ in the form
\be
A_T  =  A + 4A_{XX} - \tsfrac1{36}|A|^2 A  + \i (|A|^2)_X   A  / \mu.
\label{eq:ampi0}
 \ee
This is the usual Ginzburg--Landau equation for $A$, but with an 
additional nonlinear derivative term arising from the damping. Extensions 
to the Ginzburg--Landau equation including terms such as $\i (|A|^2)_X A$ 
are well known~\cite{doe,hoy}, but the different scalings 
involved here have the consequence that no further terms need be included 
to ensure that all terms of a given asymptotic order are present: with our 
scalings, (\ref{eq:ampi0}) is complete. Note that in the limit $\mu \to 
\infty$,  we recover the usual Ginzburg--Landau equation for $A$.

\begin{figure}
\begin{center}
\includegraphics[width=0.6\linewidth]{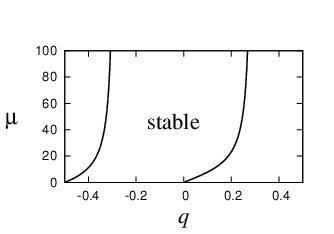}
\end{center}

\caption{Stability boundaries of rolls according to (\ref{eq:QM}), in 
$(q,\mu)$-space, where the damping coefficient $\nu=\ep\mu$. Rolls exist 
for $-1/2 < q < 1/2$ and are stable in the region between the two curves, 
where indicated. In the limit $\mu\to0$, the stable rolls are those with 
$q<0$. In the limit as $\mu\to\infty$, the usual Eckhaus stability 
boundary is recovered, so that rolls are stable for 
$-(1/12)^{1/2}<q<(1/12)^{1/2}$.}

\label{fig:margin}

\end{figure}

It is straightforward to analyse the stability of patterns in 
(\ref{eq:ampi0}) -- see Mancebo and Vega~\cite{man}. Steady patterns with 
$A = a_0 \exp{\i q X}$ exist for $q^2 < q_c^2=1/4$, with the real 
amplitude $a_0=6(1-4q^2)^{1/2}$. The stability of these solutions may be
determined by writing $A = (a_0 + a(X,T))\exp{\i q X}$ and linearizing in 
the perturbation $a$, to give
 \be
a_T = -\tsfrac{1}{36}
a_0^2 (a + a^*) + 8 \i q a_X +4a_{XX} + \i a_0^2 (a_X + a^*_X)/\mu.
\label{eq:aeq}
 \ee
After writing $a = b + \i c$, and separating (\ref{eq:aeq}) into real and 
imaginary parts, we find that $b$ and $c$ obey
 \begin{eqnarray}
b_T &=& - \tsfrac{1}{18} a_0^2 b - 8 q c_X + 4b_{XX}, \\
c_T &=& 8 q b_X + 4c_{XX} + 2 a_0^2 b_X/\mu. 
 \end{eqnarray}
Then by supposing $b$ and $c$ each to be proportional to $\exp({\lambda T 
+ \i l X})$, we find that the growth rate $\lambda$ satisfies
 \be
\lambda^2 + (\tsfrac{1}{18}a_0^2 + 8l^2) \lambda +
16 l^4 + \tsfrac{2}{9} a_0^2 l^2 - 64 q^2 l^2 - 16 q l^2 a_0^2/\mu = 0;
  \label{eq:lambdaquad}
\ee
only monotonic instability is possible, and there can be instability if 
and only if
 \be
l^2(72 l^2 + a_0^2 - 288 q^2 - 72 q a_0^2/\mu)<0.
 \label{eq:clear}
 \ee
It is clear from (\ref{eq:clear}) that the most ``dangerous'' modes are 
those with small $l$, so the pattern is {\em stable} if 
 \be
(1 - 12 q^2)  > 72 q(1-4q^2)/\mu. \label{eq:QM}
 \ee
The corresponding stability boundaries are shown in 
Figure~\ref{fig:margin}.

In the limit $\mu \to \infty$, the usual Eckhaus stability condition (that 
rolls are stable for $q^2<q_e^2 \equiv\tsfrac13q_c^2=\tsfrac1{12}$) is 
recovered from (\ref{eq:QM}). As $\mu$ is decreased, however, an asymmetry 
develops in the stability region: patterns are stable for 
$f(\mu)<q<g(\mu)$, where $f(\mu)\in[-\tsfrac12,-q_e)$ and 
$g(\mu)\in[0,q_e)$ are each increasing functions of $\mu$.  When $\mu$ is 
small, the stability condition is roughly $q(1 -4q^2)<0$, so that patterns 
with $q > 0$ are unstable and patterns with $q<0$ are stable. A 
corresponding conclusion is reached by Tribelsky (whose results are 
plotted in his Figure~1(a)). If we compare the results of this analysis 
with the numerical stability calculations described above in 
Section~\ref{sec:3}, we see that the stability boundaries in 
Figure~\ref{fig:margin} correspond, roughly, to those in 
Figure~\ref{fig:ep0.1} above $\nu\approx0.2$ (for $\ep=0.1$) and above 
$\nu\approx0.07$ (for $\ep=0.05$).

\begin{figure}
\begin{center}
\includegraphics[width=0.45\linewidth]{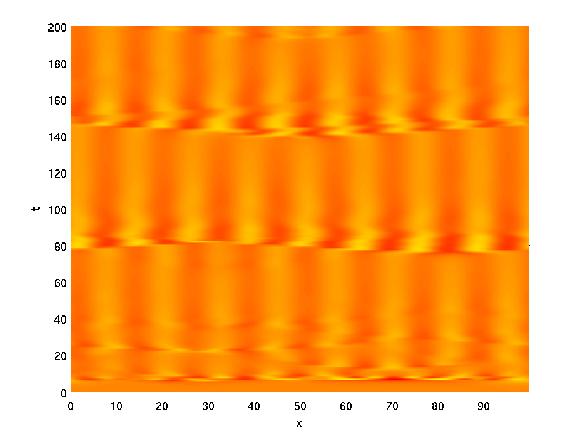}
\includegraphics[width=0.45\linewidth]{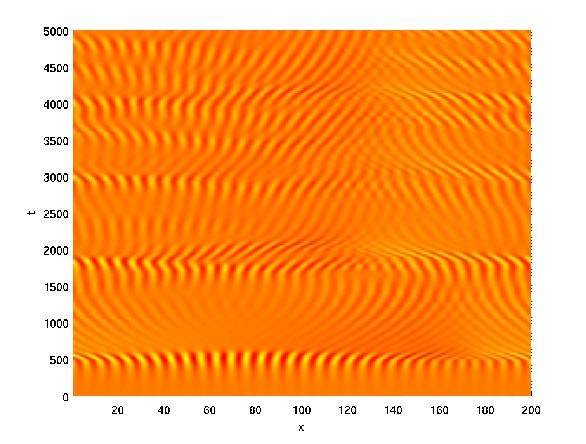}
\end{center}
\caption{Space--time plots of numerical solutions of (\ref{eq:ampi0}), for
  $\mu=0.1$ (left, real part of $A$ plotted), and the undamped
  Nikolaevskiy equation  (\ref{nik})  with $r=0.01$ (right).}
\label{fig:compare}
\end{figure}

The above analysis shows that for small $\mu$, patterns with $q<0$ are
stable in (\ref{eq:ampi0}). However, numerical simulations of
(\ref{eq:ampi0}) in a 
periodic domain, started from a random, small amplitude initial
condition do not always lead to stable patterns. A typical simulation,
for $\mu=0.1$, in a domain of size 100, is shown in
Figure~\ref{fig:compare} (left). The solution is
dominated by a state with 8 rolls in the domain, with a wavenumber 
$q = -8 \times 2\pi/100 \approx -0.5026$, which is just {\em outside} the
range of wavenumbers for which steady patterns exist. This mode must
therefore decay, until a bursting event occurs, involving other
wavenumbers, from which the  8-roll state again emerges. 
Simulations with different domain sizes show that (\ref{eq:ampi0})
seems to exhibit a preference for modes near to the left-hand limit of
the region of existence of steady states, $q=-1/2$. 
Note that this behavior is shared by the damped Nikolaevskiy equation
(\ref{eq:damp}), as shown in Figure~\ref{fig:ep0.1}. Another 
qualitative similarity between the behavior of (\ref{eq:ampi0}) and the
Nikolaevskiy equation is that both exhibit irregular bursting
behavior (albeit with rather different detailed structures). 
For comparison, a simulation of the undamped
Nikolaevskiy equation  (\ref{nik})  in a domain of size 200, for
$r=0.01$,  is shown in Figure~\ref{fig:compare} (right).

In the limit $\mu \to 0$, it is clear from (\ref{eq:lambdaquad}) that
the eigenvalues $\lambda$ become large, of order $\mu^{-1/2}$. 
This indicates that a different scaling is required in which the
damping $\nu$ is smaller than $O(\ep)$ and the growth rate of the
instability is greater than $O(\ep^2)$; this next scaling is
considered in the following section.

\subsection{Intermediate damping: $s=3/2$}
\label{sec:inter}

The analysis above corresponds to damping that is sufficiently strong to 
enslave the large-scale mode to gradients of the pattern amplitude. More 
subtle effects of damping can be explored if it is rather weaker: if 
$s=3/2$ in (\ref{eq:damp}). Our starting point is now the general roll 
solution of (\ref{eq:damp}) with $\nu=\ep^{3/2}\mu$, which may be written 
in the form
 \[
u=\ep a_0\e^{(1+\ep q)\i x}+\mbox{c.c.}+O(\ep^2),
 \] 
where $a_0=6(1-4q^2)^{1/2}$. 

A proper treatment of the stability of rolls requires consideration of the 
evolution of both amplitude and phase of the rolls, together with a 
large-scale mode. All three modes couple together, and their {\em 
relative} scalings, together with appropriate length and time scales for 
their evolution, are chosen below to enable a balance in the linearized 
perturbation problem; this is essentially the balance considered 
in~\cite{mat}.  However, by making appropriate {\em absolute} scalings for 
the three perturbation quantities, we are able to extend the linear 
results in~\cite{mat} and accommodate the first {\em nonlinear} term in 
the evolution equations that follow. We shall see below that this enables 
us to describe the sub- or supercritical nature of the onset of 
instability of the rolls. It turns out that the appropriate scaling for 
the three perturbation quantities is accomplished by writing
 \beq
u=\ep
(a_0+\ep^{1/2}a(X,T))\e^{\i x}\e^{\i 
q\ep x}\e^{\i\ep^{1/4}\phi(X,T)}+\mbox{c.c.}
+\cdots+\ep^{7/4}f(X,T)+\cdots,
\label{eq:bigu}
 \eeq
where now
 \[
X=\ep^{3/4}x,\qquad
T=\ep^{3/2}t. 
 \]
Note that in order to simplify notation we use the same symbols ($X$
and $T$) in different sections to represent {\em different} length  
and time scales. Notation within any one section is consistent, so no 
confusion should arise.
 
Then after a systematic substitution of (\ref{eq:bigu}) in (\ref{eq:damp})
and a consideration of terms at successive powers of $\ep^{1/4}$
we eventually find 
\bea
\prt{\phi}{T}&=&4\prtt{\phi}{X}-f,\label{eq:phifaphi}\\
\prt{f}{T}&=&\prtt{f}{X}-\mu f-2a_0\prt{a}{X},
\label{eq:phifaf}\\
\prt{a}{T}&=&4\prtt{a}{X}-4a_0\left(\prt{\phi}{X}\right)^2-8qa_0\prt{\phi}{X}.
\label{eq:phifaa}
 \eea 
The linear terms in this equation may be identified with the linear 
system considered in~\cite{mat}, with the addition here of the new damping 
term $-\mu f$ in (\ref{eq:phifaf}) (the quantities $b$, $c$ and $f$ 
in~\cite{mat} correspond, respectively, to our $a$, $a_0\phi$ and $f$); 
the nonlinear term was not computed in~\cite{mat}. These three equations 
may then be reduced to the nonlinear phase equation
 \beq
\left(\prt{}{T}-4\prtt{}{X}\right)^2
\left(\prt{}{T}-\prtt{}{X}+\mu\right)\phi=-16a_0^2\left(q+\prt{\phi}{X}\right)
\prtt{\phi}{X}.
\label{eq:phiq01}
\eeq

In analysing (\ref{eq:phiq01}), it is useful to consider first the 
linearized problem, for which the right-hand side is simply 
$-16a_0^2q\phi_{XX}$. This term represents the coupling between $\phi$, 
$a$ and $f$: in its absence, no instability can result in 
(\ref{eq:phiq01}). Thus in Sections~\ref{sec:5.2.1} and~\ref{sec:5.2.2} 
below we assume $a_0^2q=O(1)$. However, we should note that the analysis 
described below must be reconsidered when $|a_0^2q|$ is {\em small} 
(cf.~\cite{triv}) -- we see immediately that there are two circumstances 
in which a separate consideration is warranted ($a_0\ll1$ and $|q|\ll1$); 
we shall return to this important point below, in Section~\ref{sec:4.3}.

\subsubsection{No damping}
\label{sec:5.2.1}

We begin by summarizing relevant results in the absence of damping. When 
$\mu=0$ we recover from (\ref{eq:phiq01}) the secondary stability results 
obtained by Tribelsky and Velarde~\cite{triv} and elsewhere by 
ourselves~\cite{mat}: there is monotonic instability of the rolls with 
$q>0$, and such rolls are unstable to disturbances with wavenumbers $l$ in 
the range $0<|l|<l_{cm}\equiv(a_0^2q)^{1/4}$; there is oscillatory 
instability for $q<0$, and here the instability strikes for 
$0<|l|<l_{co}\equiv(-2a_0^2q/25)^{1/4}$. Thus all rolls are unstable at 
onset in the undamped problem.

It is instructive to extend the linearized analysis by considering the 
weakly nonlinear development of these instabilities, according to 
(\ref{eq:phiq01}). We shall focus on computing the coefficients of the 
nonlinear terms in Landau equations for the amplitudes of the 
disturbances; the signs of the real parts of these coefficients will 
indicate the direction of bifurcation to the perturbed state. We consider 
first the monotonic instability for $q>0$, and examine the evolution of a 
disturbance whose wavenumber satisfies $|l-l_{cm}|=O(\delta)$, where 
$0<\delta\ll1$. Expanding
 \be
\phi=\delta\phi_1+\delta^2\phi_2+\cdots,
 \label{eq:phiex}
 \ee
where 
\[
\phi_1=A(T_2)\sin lX
\]
and $T_2=\delta^2 T$, we find at $O(\delta^3)$ that
\[
\deriv{A}{T_2}=\frac{a_0^2}{180q}A^3.
 \]
Thus the bifurcation is always {\em subcritical} and hence leads to the 
observed explosive growth of the instability in simulations of the 
undamped Nikolaevskiy equation~\cite{mat,triv}. For the oscillatory 
instability for $q<0$, we instead suppose that $|l-l_{co}|=O(\delta)$; 
again expanding $\phi$ as in (\ref{eq:phiex}), but now with
 \[
\phi_1=\e^{\i\omega T}
\left(B(T_2)\e^{\i lX}+C(T_2)\e^{-\i lX}\right)+\mbox{c.c.},
\]
where $\omega=(-48a_0^2q/25)^{1/2}$, we find at $O(\delta^3)$ the Landau
equations
\bea
\frac{-q}{a_0^2B}\deriv{B}{T_2}&=&
\left(-\frac{740}{72639}+\frac{55\sqrt{6}}{8071}\i\right)|B|^2
+
\left(-\frac{40}{1197}-\frac{10\sqrt{6}}{399}\i\right)|C|^2,\\
\frac{-q}{a_0^2C}\deriv{C}{T_2}&=&
\left(-\frac{740}{72639}+\frac{55\sqrt{6}}{8071}\i\right)|C|^2
+
\left(-\frac{40}{1197}-\frac{10\sqrt{6}}{399}\i\right)|B|^2.
\eea
Since $q<0$, both travelling waves and standing waves branch supercritically,
with the travelling wave stable. In practice we expect the explosive 
development of the monotonic instability to dominate the smooth onset of 
the oscillatory instability in simulations of the undamped Nikolaevskiy 
equation, except under extremely controlled (and contrived) conditions.

\subsubsection{Nonzero damping}
\label{sec:5.2.2}

\begin{figure}
\begin{center}
\includegraphics[width=0.6\linewidth]{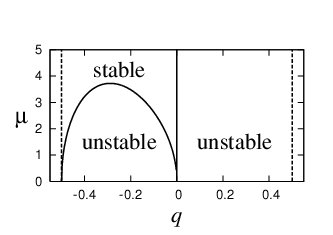}
\label{fig:inter}
\end{center}

\caption{Stability boundaries of rolls, based on analysis of the case of 
intermediate damping, $\nu=\ep^{3/2}\mu$. Rolls exist between the dashed 
lines and are stable or unstable as indicated. The solid vertical straight 
line represents the onset of a monotonic bifurcation and the solid curve 
represent a Hopf bifurcation. }

\label{fig:interm}

\end{figure}

For $\mu>0$, we again begin by examining the linearized behavior of 
(\ref{eq:phiq01}).  A monotonic instability again occurs for all rolls 
with $q>0$, with the stability margin being given by $l_{cm}^4+\mu 
l_{cm}^2-qa_0^2=0$. It thus follows that
 \[
 l^4_{cm}(\mu)=qa_0^2-\mu l_{cm}^2<qa_0^2=l^4_{cm}(0),
 \] 
and hence the rolls are unstable to a smaller range of disturbance 
wavenumbers than in the undamped case. An oscillatory instability arises 
for rolls with $q<q_0$, where $q_0=-\mu^2/(2a_0^2)$; such rolls are 
unstable to disturbances with $0<|l|<l_{co}$, where now 
$5l_{co}^2=-\mu+(-2qa_0^2)^{1/2}$. We may rewrite the condition $q<q_0$ 
for instability as
 \beq
72 q (1-4 q^2) < - \mu^2 .
\label{eq:hopfcubic}
 \eeq 
Then since the cubic function in (\ref{eq:hopfcubic}) has a minimum value 
of $-8\sqrt 3$ at $q=-\sqrt 3 / 6$, this oscillatory instability can occur 
only for $\mu^2 < 8\sqrt 3$, that is, for sufficiently weak damping (this 
is consistent with no oscillatory instability being observed in the 
previous scaling of stronger damping in Section~\ref{sec:strong}). In the 
limit of small $\mu$ the stability boundaries in (\ref{eq:hopfcubic}) 
approach $q=0$ and $q=-1/2$, so that all rolls with $q<0$ are predicted 
to be unstable to an oscillatory disturbance, just as in the undamped 
case.

The stability boundaries for this scaling are illustrated in 
Figure~\ref{fig:interm}. Note that the curved line of the Hopf bifurcation 
is consistent with the numerical stability results for the full damped 
Nikolaevskiy equation (\ref{eq:damp}) shown in Figure~\ref{fig:ep0.1} 
(except, of course, that the analysis above does not capture the 
finite-wavelength instability denoted by ``\textit{f}'' in 
Figure~\ref{fig:ep0.1}). Figure~\ref{fig:interm} suggests that two regions 
of stable rolls, near $q=0$ and $q=-1/2$, extend right down to zero 
damping.  But in each of these cusped regions, 
$|a_0^2q|=|6q(1-4q^2)^{1/2}|\ll1$, and we recall that the present scaling 
is not valid in this limit. Thus we should disregard the two cusps in 
Figure~\ref{fig:interm} at small damping and instead examine more closely 
the stability of rolls in this limit; to resolve the correct behavior of 
the stability boundary at small damping, we consider in the next section a 
further (and final) scaling for $\nu$.

\subsection{Weak damping: $s=2$}
\label{sec:4.3}

The final small-$\ep$ asymptotic description of the roll stability 
boundary may be described by setting $s=2$, so that $\nu=\ep^2\mu$. In 
this case, the damping rate of the mean mode is of the same order as the 
growth rate of the pattern-forming mode. There are two cases to consider
where the scaling of the previous section breaks down, corresponding to 
$|q|\ll1$ and $a_0\ll1$.

\subsubsection{Central corner ($|q|\ll1$)}
\label{sec:5.3.1}

As Tribelsky and Velarde~\cite{triv} have pointed out for the case of no 
damping, the scalings of Section~\ref{sec:inter} break down when $q$ is 
small. To resolve the small-$q$ behavior, we adopt Tribelsky and 
Velarde's scalings (but in an amplitude-equation framework) and 
consider
 \beq
u\sim\ep
(a_0+\ep^2a(X,T))\e^{\i x}\e^{\i\ep^2 q x}\e^{\i\ep\phi(X,T)}+\mbox{c.c.}
+\cdots+\ep^3f(X,T)+\cdots,
\label{eq:bigutv}
\eeq
where now $X=\ep x$ and $T=\ep^2 t$ and $a_0=6$. Note that
whereas (\ref{eq:bigu}) concerns basic roll wavenumbers $k=1+O(\ep)$,
(\ref{eq:bigutv}), by contrast, concerns the much narrow wavenumber band
$k=1+O(\ep^2)$. We find, after a substitution of (\ref{eq:bigutv}) in
(\ref{eq:damp}),
\bea
\prt{\phi}{T}&=&4\prtt{\phi}{X}-f,\\
\prt{f}{T}&=&\prtt{f}{X}-\mu f-2a_0\prt{a}{X},\\
\prt{a}{T}&=&4\prtt{a}{X}-\frac{1}{18}a_0^2a+
a_0\left(\frac{11}{54}a_0^2-8q+12\prtt{}{X}\right)\prt{\phi}{X}
-4a_0\left(\prt{\phi}{X}\right)^2-a_0\prt{f}{X}.
\eea
These three equations may then be reduced to the single nonlinear phase 
equation
\bea
\left(\prt{}{T}-4\prtt{}{X}\right)
\left(\prt{}{T}-4\prtt{}{X}+\frac1{18}a_0^2\right)
\left(\prt{}{T}-\prtt{}{X}+\mu\right)\phi=\nonumber\\
2a_0^2\left(\frac{11}{54}a_0^2-8q+\prt{}{T}+8\prtt{}{X}\right)\prtt{\phi}{X}
-16a_0^2\prt{\phi}{X}\prtt{\phi}{X}.
\label{eq:dampq}
\eea

\begin{figure}
\begin{center}
\includegraphics[width=0.6\linewidth]{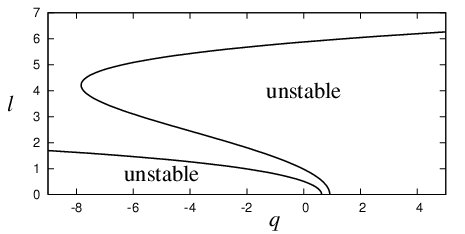}
\end{center}

\caption{Stability boundaries of small-$q$ rolls, according to 
(\ref{eq:dampq}), in the undamped case $\mu=0$ (cf.~\cite{triv}). To the 
right is the monotonic stability boundary for rolls; to the left is the 
oscillatory stability boundary.}

\label{fig:mu0}

\end{figure}

We now consider the linearized version of (\ref{eq:dampq}) and examine the 
dispersion relation for infinitesimal disturbances proportional to 
$\exp(\sigma T+\i lX)$. It is helpful first to recall the stability 
results of Tribelsky and Velarde~\cite{triv} for the case $\mu=0$, which 
are summarized in Figure~\ref{fig:mu0}, and which are a special case of 
the more general analysis to be presented below. As can be seen in the 
figure, all rolls are unstable for $\mu=0$.

\begin{figure}
\begin{center}
\includegraphics[width=0.6\linewidth]{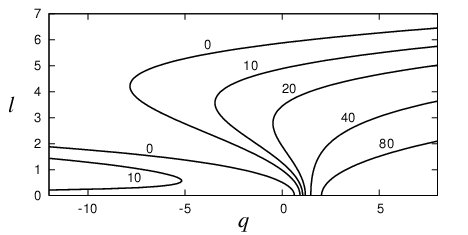}
\end{center}

\caption{Stability boundaries of rolls with damping coefficient 
$\nu=\ep^2\mu$, according to (\ref{eq:dampq}), for values of $\mu$ as 
indicated. To the right are the monotonic stability boundaries; to the 
left are the oscillatory stability boundaries. As the damping coefficient 
increases, the rolls become less unstable. For sufficiently large $\mu$, a 
stable band of wavenumbers emerges.}

\label{fig:dampq}

\end{figure}

For general values of $\mu$, rolls undergo a monotonic instability 
($\sigma=0$) when $q=q_m(l,\mu)$, where
 \[
72 q_m(l,\mu) = 2l^4 + (2\mu-71)l^2 + 66 + \mu. 
 \]
Since $q_m(l,\mu)>q_m(l,0)$, it follows that the stability margin tends 
to move to the {\em right} as the damping is increased, tending to 
stabilize the rolls (see Figure~\ref{fig:dampq}). Rolls undergo an 
oscillatory instability ($\sigma=\i\omega$) when
 \beq
q=-\frac{\mu(2+\mu)}{288}l^{-2}
-\left(\frac{\mu^2}{72}+\frac{3\mu}{16}-\frac{91}{144}\right)
-\left(\frac{5\mu}{36}+\frac{677}{288}\right)l^2
-\frac{25}{72}l^4,
\label{eq:Qosc}
\eeq
with onset frequency given by 
 \[ 
\omega^2=24l^4+2(4\mu+41)l^2+2\mu.
 \] 
The oscillatory stability boundary is plotted in Figure~\ref{fig:dampq} 
for various values of $\mu$. Damping is seen to shift the oscillatory 
stability curve to the left, again stabilizing the rolls. A further effect 
of nonzero damping is to stabilize rolls to oscillatory disturbances in 
the limit that the perturbation wavenumber $l\to0$.

\begin{figure}
\begin{center}
\includegraphics[width=0.6\linewidth]{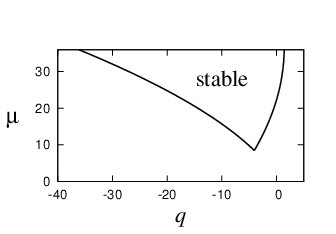}
\end{center}

\caption{Stability boundary for rolls with wavenumber $1+\ep^2q$, 
according to (\ref{eq:dampq}), for weak damping, with $\nu=\ep^2\mu$. 
Rolls are stable inside the wedge. At the right-hand boundary they become 
susceptible to a monotonic instability, at the left-hand boundary an 
oscillatory instability. The wedge terminates at the point 
$(\mu,q)\approx(8.445,-4.049)$.}

\label{fig:cusp}

\end{figure}

As can be seen from Figure~\ref{fig:dampq}, with increasing damping the 
two stability boundaries separate and eventually part, allowing a small 
band of stable rolls. We find that stable rolls exist for 
$\mu>\mu_c\approx8.445$ (the corresponding critical value of $\nu$ is 
$\nu_c=\mu_c\ep^2$); thus, for a given value of the supercriticality 
parameter, there is a threshold value of the damping coefficient to allow 
stable rolls near $k=k_c$, as previously shown by Tribelsky~\cite{try06}. 
Figure~\ref{fig:cusp} shows the region of stable rolls in $(\mu,q)$-space 
(thus the apparent small-$q$ cusp in Figure~\ref{fig:interm}, which 
appears to extend down to zero damping, is in fact a wedge terminating at 
finite damping).

\begin{figure}
\begin{center}
\includegraphics[width=0.6\linewidth]{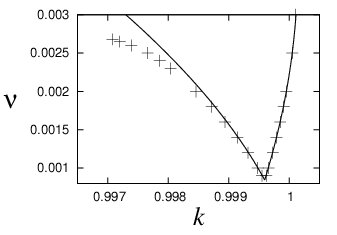}
\end{center}

\caption{Stability boundary for rolls in (\ref{eq:damp}), computed 
numerically for $\ep=0.01$, near the central wedge region corresponding to 
Figure~\ref{fig:cusp}. Crosses indicate numerical results and the solid 
line corresponds to our analytical results (cf.~Figure~\ref{fig:cusp}). 
Note that for this value of $\ep$ (and presumably also for even smaller 
values) the numerical stability boundary contains this central wedge 
rather than a section corresponding to that marked ``\textit{f}'' in 
Figure~\ref{fig:ep0.1}.}

\label{fig:ep0.0001}

\end{figure}

The results above seem at odds with the numerical results presented in 
Figure~\ref{fig:ep0.1}, since there is no central wedge in the numerical 
stability boundary. However, it turns out that for the values of $\ep$ 
used in Figure~\ref{fig:ep0.1}, the small-$q$ wedge is masked by an 
additional finite-wavenumber instability. For a smaller value of $\ep$, 
the theory in this section does indeed correctly predict the shape of the 
stability boundary, as shown in Figure~\ref{fig:ep0.0001} for $\ep=0.01$. 
(It becomes increasingly challenging to compute the entire stability 
boundary numerically, so only the relevant part is presented for this 
value of $\ep$.) Rolls are stable in the wedge down to 
$\nu\approx8.445\ep^2$, and the last rolls to destabilize have wavenumber 
approximately $1-4.049\ep^2$. These results agree with corresponding 
results of Tribelsky~\cite{try06}.

\subsubsection{Left-hand corner ($a_0\ll1$)}
\label{sec:smalla0}

We now examine the second case in which the asymptotic results of 
Section~\ref{sec:inter} break down: $a_0\ll1$. This regime proves to 
include the last rolls to become destabilized as the damping is reduced.
We thus investigate the 
stability of rolls close to the marginal stability boundary.  Recall from 
Section~\ref{sec:inter} and in particular Figure~\ref{fig:interm} that we 
expect stable rolls to persist to small values of the damping near the 
left-hand side of the marginal stability curve.

We introduce the notation $k=k_m^-,k_m^+$ for the left- and right-hand 
marginal stability boundaries, respectively (that is, for given values of 
$r$ and $\nu$, rolls exist for wavenumbers $k$ in the range 
$k_m^-<k<k_m^+$). We shall begin by considering both left- and right-hand 
parts of the marginal stability curve, since both lead to $a_0\ll1$. Our 
analysis will then show that only the left-hand part of the marginal curve 
is relevant, being capable of supporting stable rolls. We find, from 
substitution in the equation 
$k_m^2[\ep^2-(1-k_m^2)^2]-\ep^2\mu(1-k_m^2)^2=0$ for the marginal curve, 
that
 \[
k_m^{\pm}=1\pm\tsfrac12\ep-\tsfrac18\ep^2\pm(\tsfrac{1}{16}-\tsfrac14\mu)\ep^3
+\cdots.
 \] 
It turns out that the correct scaling to resolve the stability boundary 
of the rolls is obtained by setting $|k-k_m^\pm|=O(\ep^2)$, in which case 
$|\bar{u}|=O(\ep^{3/2})$, with the relevant space and time scales for the 
evolution of perturbations being $X=\ep x$ and $T=\ep^2t$. We consider 
here only the linearized secondary stability problem and we are thus at 
liberty to scale arbitrarily the absolute sizes of the three perturbation 
quantities, keeping their relative sizes fixed. Thus we suppose that
 \beq
u\sim\ep^{3/2}
(a_0+a(X,T))\e^{\i k x}\e^{\i\phi(X,T)}+\mbox{c.c.}
+\cdots+\ep^2f(X,T)+\cdots,
\label{eq:smalla0exp}
\eeq
where the wavenumber of the rolls under consideration is just inside
the marginal curve, so that
 \[
k=k_m^{\pm}\mp\ep^2\kappa.
 \]

A weakly nonlinear calculation of the roll solution itself shows that the 
amplitude and wavenumber are related by 
 \be
 a_0^2=144\kappa.
 \label{eq:a0k}
 \ee
Then a consideration of the terms in (\ref{eq:damp}) which are linear in 
the perturbation quantities $\phi$, $a$ and $f$ shows that their evolution 
is governed by
 \bea
a_0\prt{\phi}{T}&=&4a_0\prtt{\phi}{X}-a_0f\pm4\prt{a}{X},\label{eq:sa0p}\\
\prt{f}{T}&=&\prtt{f}{X}-\mu f-2a_0\prt{a}{X}, \label{eq:sa0f}\\
\prt{a}{T}&=&4\prtt{a}{X}\mp4a_0\prt{\phi}{X}.
\label{eq:sa0a}
 \eea
These three equations may then be combined to yield a single linearized 
phase equation in the form
\beq
\left(\prt{}{T}-\prtt{}{X}+\mu\right)
\left[\left(\prt{}{T}-4\prtt{}{X}\right)^2+16\prtt{}{X}\right]\phi=
\mp8a_0^2\prtt{\phi}{X}.
\label{eq:smalla0stab}
\eeq
For modes proportional to $\exp(\sigma T+\i l X)$, this gives the 
dispersion relation
\beq
\left(\sigma+l^2+\mu\right)
\left[\left(\sigma+4l^2\right)^2-16l^2\right]= \pm8a_0^2l^2.
\label{eq:smalla0sigma}
\eeq

In the limit $a_0\to0$, (\ref{eq:smalla0sigma}) gives the growth rates 
$\sigma= -(\mu+l^2)<0$, $\sigma=4l(1-l)$ and $\sigma=4l(-1-l)$, and hence 
all rolls are unstable (to a monotonic disturbance) sufficiently close to 
the marginal curve. We gain some insight into the possible existence of 
{\em stable} rolls {\em near} the marginal curve by next considering the limit 
of small perturbation wavenumbers, $l\to0$. For fixed $a_0$ and $\mu$, in 
the limit $l\to0$, the leading-order balance in (\ref{eq:smalla0sigma}) is
 \be
 \sigma^2\sim 8l^2(2\pm a_0^2/\mu). 
 \label{eq:los} 
 \ee
Thus near $k=k_m^+$ rolls are unstable, to monotonic disturbances, since 
$(2+a_0^2/\mu)>0$.  No further analysis of this case is necessary: there 
is no pocket of stable rolls possible near the right-hand marginal 
stability curve; this conclusion is consistent with 
Figure~\ref{fig:ep0.1}.

Near $k=k_m^-$, by contrast, the conclusion of monotonic instability 
follows from (\ref{eq:los}) only if $a_0^2<2\mu$ (i.e., sufficiently close 
to the marginal curve). When $a_0^2>2\mu$, (\ref{eq:los}) instead 
indicates oscillations, and we must go further in our consideration of the 
small-$l$ problem in order to determine the stability of the rolls. By 
carrying out a small-$l$ expansion of the growth rate in the form 
$\sigma=\sigma_1l+\sigma_2l^2+\cdots$, we find from 
(\ref{eq:smalla0sigma}) that $\sigma_1^2=8(2-a_0^2/\mu)$ and 
$\sigma_2=4(a_0^2/\mu^2-1)$. Thus when
 \be
2\mu<a_0^2<\mu^2,
\label{eq:mua0}
\ee
rolls are stable in the small-$l$ limit (the lower threshold for $a_0^2$ 
corresponding to a monotonic instability, the upper threshold to an 
oscillatory instability). The relation (\ref{eq:mua0}) indicates that 
rolls enjoy this region of stability only for sufficiently large damping: 
$\mu>\mu_c\equiv2$.  Note that since $2<8.445$ this cusp contains the last 
stable rolls as the damping is decreased.

In terms of the original variables in the damped Nikolaevskiy equation 
(\ref{eq:damp}), we thus predict (for small $\nu$) the cusp to be at 
$\nu\sim 2r=2\ep^2$. We may readily check these analytical results against 
the numerical secondary stability calculations of Section~\ref{sec:3}. For 
$\ep=0.1$, as in Figure~\ref{fig:ep0.1}, we thus predict the apex of the 
cusp to lie at $\nu=0.02$ (which we do indeed find numerically). 
Similarly, for $\ep=0.05$ we predict the apex to lie at $\nu=0.005$, and 
the smallest value of $\nu$ for which we have been able to compute the 
cusp is $\nu=0.006$, which provides reasonable agreement with the theory. 
(As $\ep\to0$ it becomes exceedingly difficult to compute accurately this 
part of the numerical secondary stability boundary.)

For completeness, we note that we have also considered numerically the 
full dispersion relation (\ref{eq:smalla0sigma}) for arbitrary values of 
$l$: we find that (at least for values of $a_0$ and $\mu$ up to $4$, for 
which we have carried out calculations) the stability boundary is indeed 
determined by the small-$l$ expansion.

\section{Numerical simulations}

In this section we present full numerical simulations of the damped 
Nikolaevskiy equation (\ref{eq:damp}), illustrating the transition from 
regular patterns to soft-mode turbulence as the damping is reduced. The 
simulations use periodic boundary conditions and employ a Fourier spectral 
method for the spatial discretization. The time-stepping is an explicit 
second-order Runge--Kutta form of the exponential time differencing method 
\cite{cm02} that computes the stiff linear part of the equation exactly, 
permitting $O(1)$ time steps. The size of the domain is $200\pi$, allowing a 
resolution of $0.01$ in wavenumber.

\begin{figure}
\begin{center}
\includegraphics[width=0.67\linewidth]{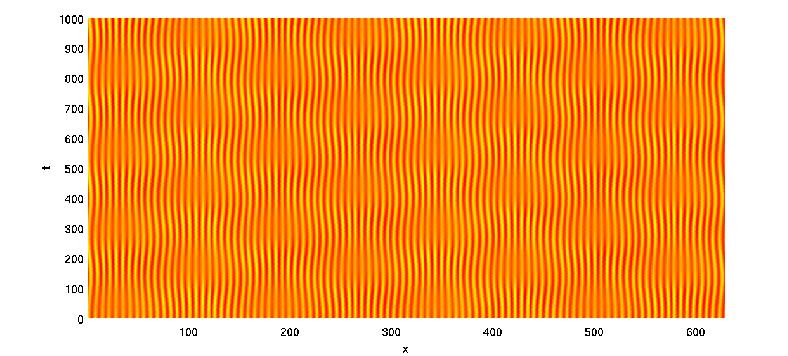}
\includegraphics[width=0.67\linewidth]{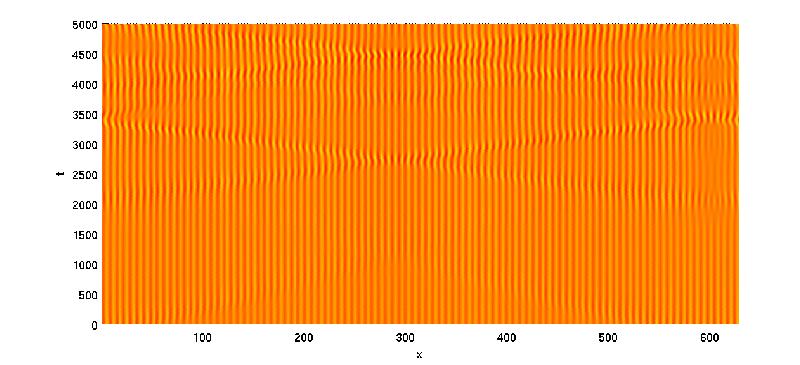}
\includegraphics[width=0.67\linewidth]{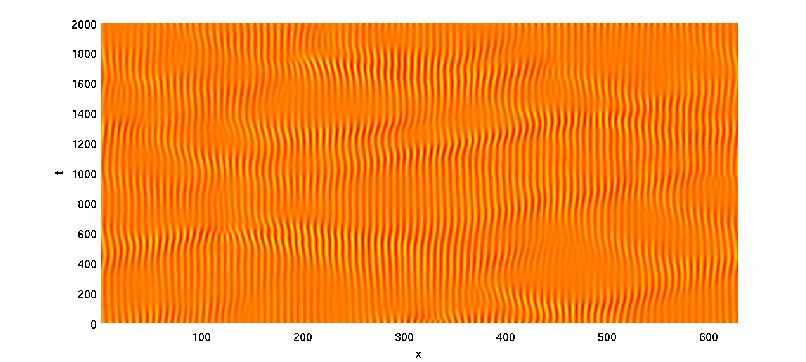}
\includegraphics[width=0.67\linewidth]{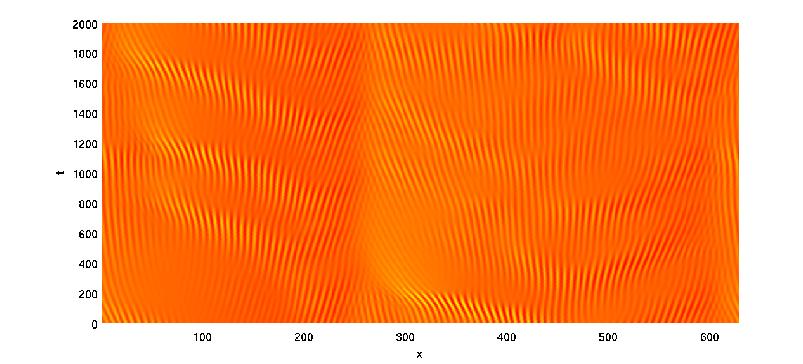}
\end{center}

\caption{Space--time plots of the numerical solution of (\ref{eq:damp}) at 
$r=0.01$. Top to bottom: $\nu=0.04$, $0.03$, $0.02$, $0$.}

\label{fig:numsim}
\end{figure}

In the simulations shown in Figure~\ref{fig:numsim}, the initial condition 
comprises rolls with wavenumber $k=0.96$, plus a small random 
perturbation.  In each case, the simulation was run for a long time to 
remove transient features, and then plotted over a subsequent relatively 
short time interval. Note that these time intervals are not the same in 
each case.  The driving parameter was fixed at $r=0.01$ and the sequence 
of simulations shows the effect of reducing $\nu$. For these parameters, 
all rolls that fit into the computational domain are predicted to be 
unstable for $\nu < \nu_c$ where $\nu_c \approx 0.057$. The bifurcation at 
$\nu = \nu_c$ is the Hopf bifurcation whose stability boundary is shown in 
Figures~\ref{fig:ep0.1} and~\ref{fig:interm}.

The simulations confirm the predicted bifurcation type, since for $\nu < 
\nu_c$ a small oscillatory modulation to the rolls grows. This bifurcation 
seems to be supercritical since the oscillations equilibrate to a small 
periodic modulation of the rolls. For $\nu=0.04$ this oscillation takes 
the form of a standing wave, as shown in Figure~\ref{fig:numsim} (top). At 
$\nu=0.03$, the solution is no longer periodic in time and shows 
occasional irregular bursts of instability. In this case, the dominant 
wavenumber is $k=0.95$, at the left-hand boundary of the roll-existence 
region (i.e., near $k=k_m^-$), consistent with the analysis of the 
preceding section and the behavior of (\ref{eq:ampi0}). Also,
counter-intuitively, the amplitude of the  
solution becomes significantly lower as the damping is decreased.  
This is because of the transition from the $O(\ep)$ scaling in the
strongly damped case to the $O(\ep^{3/2})$ scaling \cite{mat} in the
undamped, unstable regime. For 
$\nu=0.02$, the bursts are more frequent and a broad spectrum of 
wavenumbers is present in the solution. Finally, the lowest panel of 
Figure~\ref{fig:numsim} shows the case of zero damping. This is 
significantly different from the weakly damped case, showing grain 
boundaries between regions of travelling rolls.

\section{Discussion and conclusions}

The Nikolaevskiy equation is an important model of a wide range of
physical systems, including certain convection problems, phase
instabilities and transverse instabilities of fronts. It arises
naturally in systems with a finite-wavenumber
instability and a translation symmetry for the dependent
variable. However, in many applications it is likely that this
symmetry will be weakly broken. 

The damped Nikolaevskiy equation (\ref{eq:damp}) allows us to
investigate the effects of weak symmetry breaking, and describe the 
transition between the appearance of soft-mode turbulence at onset when 
$\nu=0$ and the more gradual development of complex dynamics typical of 
damped systems when the damping coefficient $\nu>0$. Our numerical 
investigation of the secondary stability problem for steady roll solutions 
shows how the instability of {\em all} roll states in the undamped case 
evolves into the more common Eckhaus scenario, whereby rolls with 
wavenumbers sufficiently close to critical are stable, when the damping is 
sufficiently great. If we fix the supercriticality parameter $r$, then it 
follows from the asymptotic results of Section~\ref{sec:smalla0} that 
there is a critical value of the damping coefficient $\nu=\nu_c$ (given by 
$\nu_c\sim2r$ when $r$ is small) below which all rolls are unstable; for 
$\nu>\nu_c$ some rolls are stable. 

However, it is more common in applications to fix parameters such as the 
damping coefficient $\nu$ and vary the supercriticality parameter $r$. 
Then the results of Section~\ref{sec:smalla0} may be interpreted as 
indicating that, at least for small values of $\nu$, there are stable 
rolls provided $2r<\nu$, i.e., close enough to onset. Thus the damped 
Nikolaevskiy equation differs fundamentally from the undamped version in 
that there is not a {\em direct} transition to soft-mode turbulence at 
onset when $\nu>0$; instead, there is a sequence of bifurcations beginning 
with some initially stable roll state. Of course, when $\nu$ is small this 
bifurcation sequence may occur {\em very close} to onset (i.e., for 
$r=O(\nu)$).

Our work complements a recent study by Tribelsky~\cite{try06}, but differs 
in some significant respects. The first and most important difference is 
our discovery and analysis of the cusp of small-amplitude stable rolls, as 
described in Section~\ref{sec:smalla0}. These rolls are the last to be 
destabilized as the damping is reduced, and so are particularly 
significant. Another difference is that our numerical study of the 
secondary stability problem has shown that rolls in the central wedge of 
apparent stability, which is predicted both by ourselves and by Tribelsky 
using asymptotic arguments, are in fact unstable to short-wave modes, 
except for exceedingly small values of $\nu$. So this central wedge (see 
Section~\ref{sec:5.3.1}) in fact forms part of the stability boundary only 
for very small damping.

Finally, we reiterate that our choice of damping term in
(\ref{eq:damp}) is by no means unique: indeed, Tribelsky~\cite{try06}
made a different choice, as discussed in Section~\ref{sec:damped}. We
conclude by noting that we have also computed numerically the
secondary stability boundaries corresponding to those shown in
Figure~\ref{fig:ep0.1}, but for Tribelsky's version of the damped
Nikolaevskiy equation, and find similar results.

\end{document}